\documentclass[a4paper]{article}
\usepackage{fullpage}
\usepackage{hyperref}
\usepackage{graphicx}
\usepackage{amsmath, amssymb, amsthm, mathtools}
\usepackage{xcolor}
\usepackage{tcolorbox}

\usepackage{bm} 
\newcommand{\tb}{\bm{t}}
\newcommand{\gb}{\bm{g}}
\newcommand{\ub}{\bm{u}}
\newcommand{\lb}{\bm{\ell}}
\newcommand{\covid}{COVID-19}

\theoremstyle{definition}

\newtheorem{definition}{Definition}

\title{Optimal Testing and Containment Strategies for Universities in Mexico amid COVID-19}

\author{
Luis Benavides-Vázquez%
\footnote{Monterrey Institute of Technology and Higher Education, Mexico, 
\href{mailto:luisbv@tec.mx}{luisbv@tec.mx}}
\and
Héctor Alonso Guzmán-Gutiérrez%
\footnote{San Luis Potosí Institute of Scientific Research and Technology, Mexico, 
\href{mailto:hector.guzman@ipicyt.edu.mx}{hector.guzman@ipicyt.edu.mx}}
\and
Jakob Jonnerby%
\footnote{Imperial College London, UK,
\href{mailto:l.jonnerby21@imperial.ac.uk}{l.jonnerby21@imperial.ac.uk}}
\and
Philip Lazos%
\footnote{IOHK, Italy, 
\href{mailto:philip.lazos@iohk.io}{philip.lazos@iohk.io}}
\and
Edwin Lock%
\footnote{University of Oxford, UK, 
\href{mailto:edwin.lock@economics.ox.ac.uk}{edwin.lock@economics.ox.ac.uk}}
\and
Francisco Marmolejo-Cossío%
\footnote{Harvard University, USA,
\href{mailto:fjmarmol@seas.harvard.edu}{fjmarmol@seas.harvard.edu}}
\and
Ninad Rajgopal%
\footnote{University of Warwick, UK, 
\href{mailto:ninad.rajgopal@warwick.ac.uk}{ninad.rajgopal@warwick.ac.uk}}
\and
José Roberto Tello-Ayala%
\footnote{San Luis Potosí Institute of Scientific Research and Technology, Mexico,
\href{mailto:ro9teaya@yahoo.com.mx}{ro9teaya@yahoo.com.mx}}
}

\date{}

\begin{document}

\maketitle

\begin{abstract}
    This work sets out a testing and containment framework developed for reopening universities in Mexico following the lockdown due to \covid{}. We treat diagnostic testing as a resource allocation problem and develop a testing allocation mechanism and practical web application to assist educational institutions in making the most of limited testing resources. In addition to the technical results and tools, we also provide a reflection on our current experience of running a pilot of our framework within the Instituto Tecnológico y de Estudios Superiores de Monterrey (ITESM), a leading private university in Mexico, as well as on our broader experience bridging research with academic policy in the Mexican context.
\end{abstract}

\begin{tcolorbox}
This manuscript is a reprint of a paper accepted at the ACM Conference on Equity and Access in Algorithms, Mechanisms, and Optimization in 2021 (EAAMO'21), and has not been updated since. For more information, contact \href{mailto:edwin.lock@economics.ox.ac.uk}{edwin.lock@economics.ox.ac.uk} or \href{mailto:fjmarmol@seas.harvard.edu}{fjmarmol@seas.harvard.edu}.
\end{tcolorbox}

\section{Introduction}
\label{sec:intro}
Schools and universities around the globe have suffered extended closures due to regional and national lockdown measures following the \covid{} pandemic. Due to the severe impact of these closures on education, mental health and social divides, it has been strongly recommended that educational institutions reopen as quickly and safely as possible to ensure that future generations are not held back further \cite{national2020reopening, sahu2020closure}. At the same time, it has become clear that comprehensive testing strategies, including asymptomatic screening, are essential to combat the spread of the virus \cite{cheng2020diagnostic, horton2020offline}.
As educational institutions around the world prepare to reintroduce in-person teaching, it is imperative that they do so safely, with the help of judicious testing and containment strategies \cite{panovska2020determining}. This is particularly challenging in low- and middle-income countries (LMICs) with severe constraints on testing resources. We present a testing and containment framework aimed at helping educational institutions in LMICs make the most of extremely scarce testing resources, and also provide an online allocation software to guide decision-making. Our methods, which are currently being piloted at the Instituto Tecnológico y de Estudios Superiores de Monterrey (ITESM)--a leading private university in Mexico--make use of the heterogeneity of university populations as well as pooled qPCR tests in order to balance two competing objectives: minimising potential viral spread (and subsequent critical cases), as well as minimising the number of healthy individuals who unnecessarily self-isolate under a given containment protocol. 

We consider the setting in which the population is divided into categories based on characteristics which include individuals' potential exposure to the virus (based on their occupation in the institution), their geographical location, and/or the potential of an infection becoming critical.
The testing strategies we propose divide a given budget of tests among the population categories. Our mechanism utilises \textit{group testing},
where samples from multiple individuals are pooled and tested with a single qPCR test. A positive test result indicates that at least one individual in the pool is infected, whereas a negative result indicates that all individuals tested are healthy. As a function of the test results, we propose simple containment mechanisms and subsequently measure the performance of different test allocations according to how many critical cases are prevented overall and how many individuals from a given category unnecessarily self-isolate. 






\paragraph{Outline.}
\label{sec:outline}
In Section \ref{sec:model}, we introduce our testing and containment mechanism, and outline the contagion model we developed to capture viral spread within heterogeneous university populations. Section \ref{sec:multi_opt} describes the family of testing and containment protocols we consider, as well as formal expressions for the multiple objectives we optimise for via computation. We provide details on our current pilot in Section \ref{sec:pilot-details}, with an emphasis on three main deliverables from our software: extrapolating relevant model parameters from university data, computing the Pareto frontier  of testing allocations, and providing a user-friendly interface for navigating and choosing testing allocations along the Pareto frontier. Section \ref{sec:simulations} provides further simulation evidence that our testing allocations provide long-term results over sustained periods of infection within a population. Section \ref{sec:reflections} provides a reflection regarding our experience bridging research and practice in the Mexican context. Finally, Section \ref{sec:future-work} elaborates on next steps for our pilot.


\paragraph{Related Work.}
\label{sec:related-work}

The use of testing resources to mitigate the spread of an infectious pathogen has been intensely debated during the current pandemic \cite{Larremoreeabd5393, Cleevely2020}. The idea of combining several samples into a single group test in order to reduce the numbers of tests required has previously been studied in a substantial body of literature in Computational Learning Theory \cite{dorfman1943detection, wolf1985born, du2000combinatorial, du2006pooling, cheraghchi2012}. Group testing has been applied against HIV and other diseases \cite{Wein1996, Mcmahan2012}; in the current pandemic, it has been verified experimentally with SARS-CoV2 samples \cite{Yelin2020Pooling, Sanghani2021}. A growing body of work has investigated the use of group testing as a possible way towards testing large parts of the population \cite{Gollier2020, Calabrese2020, Baker2021, Du2020, Augenblick2020, Mutesa2021}. Work in the testing literature has largely focused on minimising the number of tests required to fully determine precisely whom is infected in a population. Turning the problem on its head, we instead consider the problem of identifying a mechanism that maximises the benefit of a fixed (and scarce) weekly testing budget.


\section{Our Model for Testing and Containment}
\label{sec:model}
We consider a heterogeneous population of $n$ individuals partitioned into $k$ disjoint categories $C_1, \dots, C_k$ of sizes $n_1, \dots, n_k$. The partitioning is chosen to capture the heterogeneity of the population, with `similar' individuals being placed in the same category. On a school and university campus, one might partition the population into school students, undergraduates, postgraduates, academic teaching staff, and administrators.

Throughout, we assume that the educational institution in question has a limited budget of $T$ tests per time period that it wants to use optimally.\footnote{In our pilot study, the university has sufficient resources to test approximately $1-6\%$ of a campus population individually every week. For most campuses this results in a testing budget of around 20-50 tests per week.} We treat the decision of how to maximise the use of limited testing resources to the population categories as a resource allocation problem that balances two objectives: reducing the virus spread and minimising the impact of quarantining on the population. In order to achieve this, we define the following testing and containment protocols, for which we can subsequently quantify these two objectives. Our protocols approach this problem via the use of \textit{pooled} or \textit{group} tests.

\paragraph{Group Testing.} In what follows, we assume that we have access to group testing (also known as pooled testing) for \covid{}. In a group test of size $g \in \mathbb{N}$, samples from $g$ individuals are pooled into one sample and subjected to a single qPCR test. A positive result on the pooled test implies that at least one of the $g$ individuals is infected, and a negative result implies that all are healthy. We let $G \in \mathbb{N}$ denote the upper bound on feasible group sizes, which is dictated by biological and laboratory constraints. {Our partner laboratories in Mexico have validated the group testing methodology of Sanghani et al.~\cite{Sanghani2021}, which permits testing with groups up to size $10$.}

\paragraph{Testing Strategies.}
Recall that $T$ is the number of tests available to the educational institution per time period. We consider a space of testing protocols parameterised by a pair of vectors $(\tb, \gb)$, where $\tb \in \mathbb{N}^k$ and $\gb \in \{1, \ldots, G\}^k$. A \textit{testing strategy} $(\tb, \gb)$ specifies the number of tests allocated to each population category, and their group sizes: $t_i$ tests are allocated to disjoint groups of size $g_i$ from category $C_i$ uniformly at random.\footnote{Note that $g_i = 1$ for some category $C_i$ implies that the testing strategy performs individual tests for category $C_i$.} Note that this implies the constraints $g_i \leq G$ and $t_i g_i \leq n_i$, for every category $C_i$. We also impose the budget constraint $\sum t_i = T$ on any testing strategy, as we wish to maximise the use of the given testing budget. We say that a testing strategy $(\tb,\gb)$ is feasible if it satisfies these constraints.

\paragraph{Containment Protocol.}
We propose a conceptually simple containment protocol: whenever a group test is negative, everyone in the group continues to function normally. If the test is positive, everyone in the group is told to self-isolate for a given period of time. In our pilot study, individuals are quarantined for 14 days. Note that this reduces the number of tests required to test a large number of individuals, and thus has the potential of catching more infectious people. On the other hand, it might lead to healthy individuals self-isolating unnecessarily. In Section~\ref{sec:multi_opt}, we describe how testing strategies can choose a desired trade-off between these conflicting objectives for each category, by selecting appropriate group sizes at which to test individuals in each category.

\paragraph{Contagion Model.}
\label{sec:formal_details_model}
For each category, we define parameters that govern contagion in the population. Let us consider an arbitrary category $C_i$. We assume that all individuals in category $C_i$ are independently infected with probability $p_i \in [0,1]$ (and healthy with probability $q_i = 1-p_i$). Each newly infected individual in $C_i$ has probability $v_i$ of developing a ``critical'' infection. Here the scope of `criticality' is defined by the university; it might, for instance, denote all cases leading to hospitalisation or death, or it may encompass all symptomatic infections.

After performing a given testing strategy $(\tb, \gb)$, it is possible that some infected individuals are not self-isolating, as they have not been subjected to a group tests. In order to quantify the performance of different testing and containment protocols, our model assumes a single step of contagion, whereby each infected individual who is not self-isolating may infect others. Each individual in $C_i$ is assumed to be in contact with $d_{ij} \in \{0,\dots,n_j\}$ individuals from $C_j$. Moreover, each susceptible (healthy) individual in category $C_i$ is infected by an infectious neighbour in $C_j$ with transmission probability $\pi_{ij} \in [0,1]$. The parameters $d_{ij}$ and $\pi_{i,j}$ can be interpreted as connectivity or `exposure parameters' between categories $C_i$ and $C_j$. We note that $d_{ij}$ and $\pi_{i,j}$ need not be symmetric.

\section{COVID Testing Allocation as an Optimisation Problem}
\label{sec:multi_opt}

We can now formally define the optimisation problem underlying our choice of optimal testing strategies to contain viral infections while minimising the disruption on education at the institution. Recall that our goal is to balance the two goals of reducing `critical cases' while minimising the number of unnecessarily self-isolating individuals.

\subsection{Our Objectives}
\label{sec:opt_alloc}
For a given feasible testing strategy $(\tb, \gb)$, we define our \textit{health objective} $O_H(\tb, \gb)$ as the expected number of critical cases that are prevented in our single-step contagion model when compared to no testing. For each category $C_i$, we also define the \textit{category quarantine objective} $O_{Q,i}(\tb, \gb)$, which denotes the expected number of unnecessarily self-isolating individuals in category $C_i$ under $(\tb, \gb)$. Notice that by minimising $O_{Q,i}$, a mechanism prioritises preventing unnecessary quarantining of individuals in the $i$-th category.

The health objective $O_H$ is optimised by covering more individuals in tests (especially from segments with high infection or connectivity to other segments), whereas the containment objectives $O_{Q,i}$ are optimised by reducing the number of healthy individuals that are quarantined unnecessarily as a result of a positive group test. Note that our objectives are conflicting: larger group sizes increase the reach of testing and lead to fewer untested but infected individuals, while smaller group sizes reduce the number of healthy individuals who are quarantined unnecessarily. 


\paragraph{Preventing Critical Cases.}
To completely specify our health objective $O_H$, we define the following auxiliary variables:
\begin{itemize}
    \item The probability $u_i = \frac{n_i - t_ig_i}{n_i}$ that a given individual in $C_i$ is not tested.
    \item The probability $z_i = u_i q_i + (1-u_i)q_i^{g_i}$ that an individual in $C_i$ is healthy and not quarantining before the contagion step of our model.
    \item The expected probability 
    \[\alpha_{i,j} =  \left( p_j u_j(1-\pi_{i,j}) + (1 - p_j u_j) \right)^{d_{ij}}\] 
    that an individual from $C_i$ who is healthy and not under quarantine becomes infected from untested and infected individuals in $C_j$.
    \item The expected number $f_H(\tb, \gb) = \sum_{i} n_i v_i z_i \left( 1 - \prod_j \alpha_{i,j} \right)$ of critical cases that occur in the contagion step of our model.
\end{itemize}
With this in hand, we can express the health objective as follows:

\begin{equation*}
O_H(\tb, \gb) = f_H(\bm{0},\bm{0}) - f_H(\tb, \gb)
\end{equation*}

To provide some intuition for this expression, let us focus on a given individual in $C_i$. If they are in a positive test, they quarantine and are hence not susceptible for contagion. If they are healthy and not under quarantine (which happens with probability $z_i$), they may then be infected by untested, infected individuals from any $C_j$. An individual in $C_j$ is untested with probability $u_j$, and the overall probability that an infection from $C_j$ is received from $d_{ij}$ interactions from individuals in $C_j$ is given by $\alpha_{i,j}$. As we are interested in the number of critical contagion infections we prevent, we measure the performance of $(\tb, \gb)$ relative to a testing strategy where no tests are used (given by $f_H(\bm{0},\bm{0})$). We refer to Appendix \ref{sec:appendix_health_objective} for further details on the construction of $O_H$.

\paragraph{Minimising Unnecessary Quarantine.} Although it is desirable for our testing strategies to maximise $O_H$, doing so comes at the cost of increasing the number of individuals who are told to unnecessarily quarantine. To account for this, recall that the \textit{category quarantine objective} $O_{Q,i}$ gives the expected number of unnecessary quarantines in category $C_i$ for each testing strategy. Each individual is independently healthy with probability $q_i = 1 - p_i$. This means that for a single group test of size $g_i$, the probability that the test is positive is given by $1 - q_i^{g_i}$, and the expected number of healthy individuals in a group of size $g_i$ conditioned on a positive test is $g_i - \frac{g_ip_i}{1-q_i^{g_i}}$. With this in hand, we can compute the expected number of healthy individuals that are under unnecessary quarantine after a single group test as follows: $g_i(q_i-q_i^{g_i})$. Finally, since we have $t_i$ of such group tests allocated randomly to $C_i$, we obtain the complete expression for the $i$-th quarantine objective:
\begin{equation*}
O_{Q,i}(\tb, \gb) = t_ig_i(q_i - q_i^{g_i})
\end{equation*}

\subsection{Solutions on the Pareto Frontier}
As mentioned in the previous sections, our objectives compete with each other and it follows that there no testing strategy that dominates all other strategies in all objectives. On the other hand, it is possible to rule out certain strategies that perform worse than other potential feasible strategies. The remaining solutions lie on the Pareto frontier, which is a natural solution concept in multi-objective optimisation. Defined below, the Pareto frontier, provides a set of mechanisms that maximally exemplify the trade-offs incurred in all objectives.

\begin{definition}[Pareto Dominance]
Suppose that $(\tb, \gb)$ and $(\tb ', \gb ')$ are two distinct testing strategies. We say that $(\tb, \gb)$ \emph{Pareto-dominates} $(\tb ', \gb ')$ if, and only if, the following hold:
$$
O_H(\tb, \gb) \geq O_H(\tb ', \gb ') \text{ and } O_{Q,i}(\tb, \gb) \leq O_{Q,i}(\tb ', \gb ') \ \forall i \in [k] 
$$
with one of the inequalities being strict. We denote this relation by $(\tb, \gb) \succ_P (\tb ', \gb ')$.
\end{definition}

Our main approach consists of precisely finding the family of feasible testing strategies that are not Pareto-dominated. 

\begin{definition}[Pareto Frontier]
The Pareto frontier $S_P$ consists of the set of testing strategies that are not Pareto-dominated by any other testing strategy.
\end{definition}

We compute the Pareto frontier $S_P$ for the class of testing mechanisms described above. Furthermore, we provide a policymakers with a principled and intuitive way to choose testing mechanisms from $S_P$ that fit their institutional needs. We provide more details on the computation of $S_P$ and the navigation tool in the next section, where we delve into the specifics of the current pilot we have underway.

\section{Details of the Pilot Study}
\label{sec:pilot-details}

A pilot of our methodology is currently underway at several campuses of the Instituto Tecnológico y de Estudios Superiores de Monterrey (ITESM). The pilot campuses were chosen for two reasons: they are in close proximity to university testing facilities with the capability to carry out group testing, and they are located in states that have recently exited lockdown. Therefore, students are able to return to a ``hybrid'' teaching format, in which a subset of students are allowed to return to in-person classes, and the university provides preemptive monitoring of infections via a limited testing budget per campus.

In line with our model from Section \ref{sec:model}, our partners have partitioned the population of each campus into 4 categories: faculty (teaching and research), administrative assistants, middle/high school students\footnote{Our partner institution also teaches middle and high school students at various campuses}, and undergraduate/graduate students. Health administrators from the university are already using our software to guide their decision-making in terms of how to allocate limited \covid{} tests per campus. The ITESM is a leading private university in Mexico with a combined student and staff population of over 96,000 individuals, all of whom are tested under allocations recommended by our software. Our software pipeline completes three key tasks: 
\begin{enumerate}
    \item Extrapolating key model parameters from university data.
    \item Computing the Pareto frontier of testing allocations given model parameters.
    \item Providing a user-friendly tool for navigating multiple solutions along the Pareto frontier of allocations.
\end{enumerate}


\subsection{Model Parameter Estimation}
\label{sec:parameter-estimation}
We estimate the parameters for the model using data provided by our university partners from Mexico pertaining to course records, attendance records for these courses, and information regarding the buildings visited by the individuals in the population. More specifically, our partnering university in Mexico currently maintains internal anonymised databases with the following information:
\begin{itemize}
    \item Per-individual information regarding membership in the population categories described above. 
    \item Information on taught courses (instructors, student attendance, classroom size/ventilation, classroom location).
    \item \covid{} testing results.
    \item Residual water test results.\footnote{Residual water results can alert administration of the presence of \covid{} at the building level of a given campus\cite{tran2020sars}.}
\end{itemize}

We use this information to directly estimate the connectivity parameters $d_{ij}$ of each of the categories specified. To do so we look at each individual from category $C_i$ and count the number of classroom or office interactions they have with individuals of category $C_j$ (we count repeat interactions as well). We take the average over all individuals in $C_i$ to produce $d_{ij}$. Notice that $d_{ij}$ is not necessarily integral, but we can compute the health objective $O_H$ as per the formula in Section \ref{sec:opt_alloc}. This process is repeated for all $i,j \in [k]$ (where we allow $i=j$). 

In order to estimate transmission probabilities, we implement methodology from Buonanno et al.~\cite{buonanno2020quantitative}, alongside key input from epidemiological experts from our partner university. More specifically, the epidemiologists provide us with reasonable estimates to key parameters to the model of Buonanno et al.~(such as room ventilation rates and individual inhalation/exhalation rates), which are informed by the data contained within the databases mentioned above, since they contain information on the quantity of individuals who sit in classroom at any given time of instruction. For a given $i,j \in [k]$, we let $\pi_{ij}$ be the average transmission probability from individuals of $C_j$ to individuals of $C_i$ as per our contagion model.

Baseline probabilities of infection $p_i$ are estimated using results of the previous \covid{} tests and additional residual waste water tests. Finally, although the vulnerability rate $v_i$ of a category will eventually be provided to us by epidemiological experts from our partner university, we are currently running our optimisation with $v_i = 1$ for the pilot, so that $O_H$ represents the total number of infections prevented.

\subsection{Computing the Pareto Frontier}
\label{sec:navigate_pareto_front}
Our algorithm for computing the Pareto frontier follows a straightforward approach that iteratively elicits the solutions in $S_P$. Given fixed model parameters, we first compute the number of critical cases that occur when no testing is applied ($f_H(\bm{0},\bm{0})$ as per Section~\ref{sec:opt_alloc}).

Next, initialise $S_P$ as the empty set. We iterate over all possible testing strategies $(\tb, \gb)$ that satisfy testing budget and group constraints, and evaluate performance under the health and per-category quarantine objectives $O_H$ and $O_{Q,i}$. For each testing strategy, we have three possible outcomes: If $(\tb, \gb)$ Pareto-dominates existing solutions in $S_P$, we add it to $S_P$ and remove all dominated solutions. If $(\tb, \gb)$ is incomparable to any element of $S_P$, we add it to $S_P$ without removing any other solutions. Finally, if $(\tb, \gb)$ is Pareto-dominated by an element of $S_P$, we do nothing. Once we have iterated over all testing strategies, $S_P$ contains the Pareto frontier.

\paragraph{Pruning the Frontier.}
Recall from Section \ref{sec:model} that a testing strategy $(\tb, \gb)$ is feasible if $g_i \leq G$, $\sum t_i = T$, and $t_ig_i \leq n_i$ for each category $C_i$. Although this represents a rich space of potential test allocations, our conversations with university administrators and testing personnel revealed a strong preference for only exploring a subset of potential testing allocations due to laboratory and logistical constraints. Furthermore, iterating over a reduced set of potential solutions has the added benefit of improving the running time of our algorithm, and providing a smaller resulting Pareto frontier, which in turn is more amenable to decision making. For this reason, our pilot iterates over group sizes in the range $g_i \in \{1,3,5,10\}$ for each category. 

Our conversations with university policymakers also revealed a preference for being presented with fewer solutions from the Pareto frontier to decide from at the time of picking testing allocations. In particular, it was felt that testing strategies that achieve very similar outcomes in all objectives should be treated as identical, and one representative solution be chosen for retention in the Pareto frontier. To address this, we implemented a ``bucketing'' scheme, whereby the performance of a given strategy $(\tb, \gb)$ under $O_H$ and $O_{Q,i}$ was truncated so that solutions that lie in the same range of values (the bucketing size) are bucketed together. The set of solutions returned by the algorithm contains one (randomly picked) representative per bucket, achieving the desired goal of reducing the solution space that policymakers navigate. The bucketing sizes per objective can be set as a parameter in our software. Alternatively, we also provide a tool that allows users to specify the desired number of solutions they wish to view, and determines optimal bucketing sizes to (approximately) achieve this. More details on our bucketing methods and our tool to restrict the number of solutions can be found in Appendix \ref{sec:appendix_buckets}.

\subsection{Web Application}
\label{sec:webapp}
We have developed a web application that assists university administrators in exploring the Pareto frontier of testing allocations to identify testing strategies that match their priorities. A demo of the navigation tool is available at \href{https://eaamo-demo.azurewebsites.net/}{our website} and the source code is provided at \href{https://github.com/hguzmang/eaamo}{our GitHub repository}. 
Figure~\ref{fig:webapp} in appendix \ref{sec:appendix_web} shows a demo screenshot, in which we have used model parameters similar to those extrapolated from our partner university as detailed in Section \ref{sec:parameter-estimation}. 
Pareto-dominant testing strategies are computed with the bucketing technique described in Section~\ref{sec:navigate_pareto_front}, and are made available to view through the web application. For each strategy $(\tb, \gb)$, the application displays how well it performs on each of the health and quarantine objectives. In order to identify desirable strategies, administrators can set thresholds on the expected number of people in each category unnecessarily self-isolating and the number of critical new cases, allowing them to find the desired balance between the different objectives. In general, there will be more than one strategy that satisfies the thresholds. The app shows the number of strategies, as well as their containment and health outcomes, and allows the user to iterate through and compare multiple solutions. Moreover, the application allows the user to save solutions for future reference.



\subsection{Preliminary Results}
In a preliminary run of our pilot, our algorithm explored 2739 allocations for one campus, and identified 80 solutions as lying in the Pareto frontier. After applying sensible bucketing parameters to filter out solutions with almost identical outcomes, only 20 solutions remained.
It follows that our algorithm categorically filters out 97\% of potential testing allocations without bucketing, and 99\% of solutions with our bucketing technique. This reduced solution set was well-received by the university administrators tasked with selecting a testing strategy for the campus.


\section{Simulations}
\label{sec:simulations}
\begin{figure}[ht]
    \centering
    \includegraphics[width=0.45\textwidth]{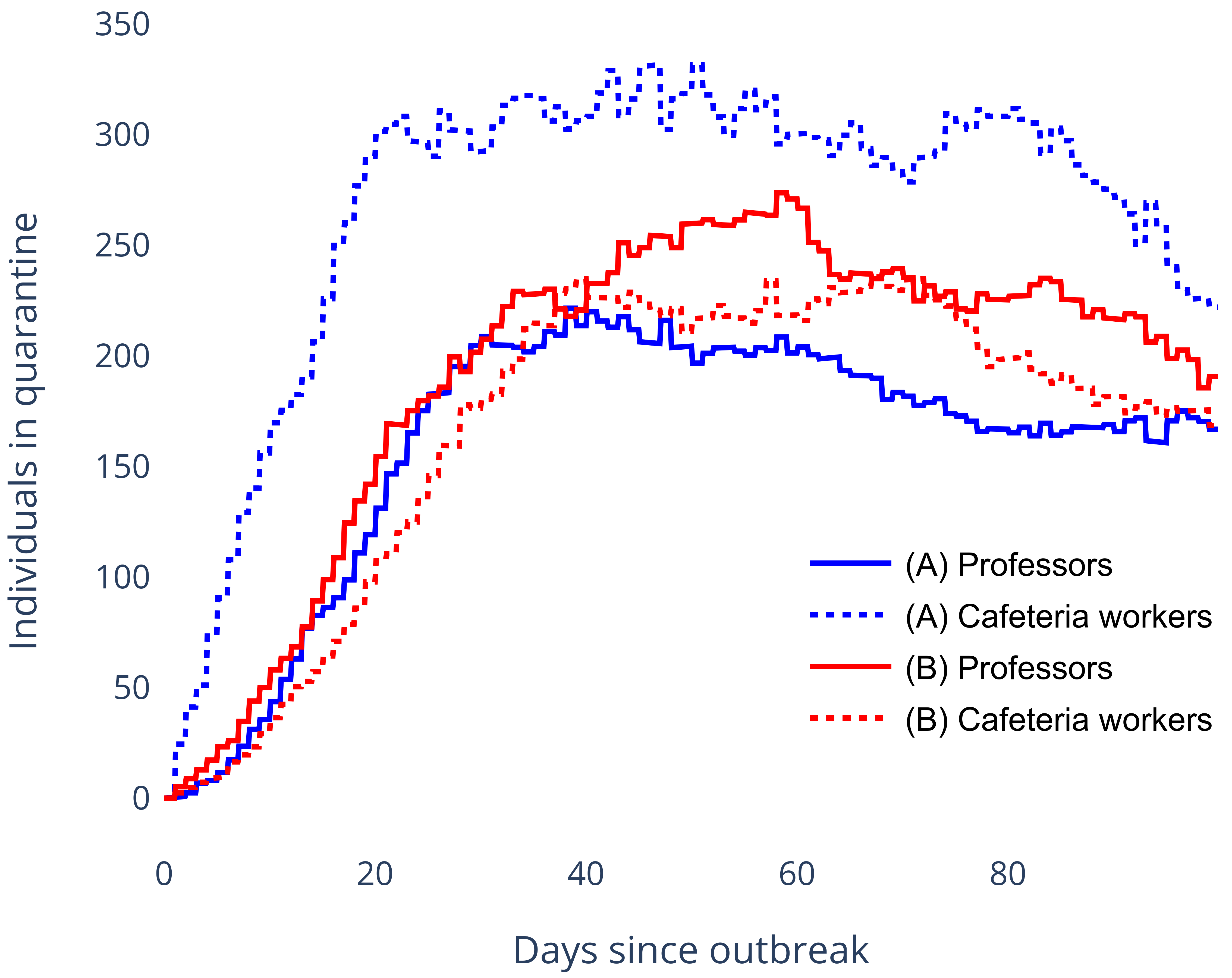}
    \caption{Number of quarantined professors (solid lines) and cafeteria workers (dotted lines) for two different allocation profiles $(A)$ and $(B)$. $(A)$ prioritises minimising professor quarantine over cafeteria workers, and in $(B)$ both have equal priority.}
    \label{fig:quarantined}
\end{figure}

We developed a network - based susceptible - infected - recovered - quarantined (SIRQ) model on a simulated university population consisting of $9,000$ students, $500$ professors, and $500$ cafeteria workers. Professors and cafeteria workers were assumed to have a much higher degree of connection than students. Using the definition of the exposure parameter $d_{ij}$ from Section \ref{sec:model} and the method for estimating it from Section \ref{sec:parameter-estimation}, we obtained
\begin{align}
   d_{ij} = 
\begin{pmatrix}
4.92 & 1.34 & 1.27 \\
24.28 & 1.44 & 1.26 \\
23.0  & 1.26  & 1.44 
\end{pmatrix},
\end{align}
with $i=1$ assigned to students, $i=2$ to cafeteria workers, and $i=3$ to professors. A simple example of how $d_{ij}$ can be used is the following: on average, a professor is exposed to 23 students ($d_{3,1}$) whereas a student is only exposed to $1.27$ professors ($d_{1,3}$). In the infection model we developed, at each time step an infected node recovers with probability $\gamma=0.0427$ and, unless it is quarantined, infects one or more of its susceptible neighbouring nodes, each with probability $\beta=0.01$. These parameters were chosen such that average number of secondary infections is $R_0\sim 3$ on the simulated network \cite{Mahase2020}. It has been shown that the testing and containment strategy we developed performs better than a random allocation of tests \cite{Jonnerby2020, Jonnerby2020a}. In these simulations, we show two different solutions on the Pareto frontier. We select allocation profiles with different values of the quarantine objectives $O_{Q,i}$. The results are shown in Figure~\ref{fig:quarantined}. This plot shows the number of individuals in quarantine on the ordinate whilst the abscissa of the graph accounts for the number of days since the outbreak started. Four curves are presented signalling two different allocation profiles, $(A)$ and $(B)$, and two different categories, professors, and cafeteria workers. In allocation profile $(A)$, represented with blue, we prioritise minimising professor quarantine over cafeteria workers and students. In allocation profile $(B)$, represented with red, we prioritise both professors and cafeteria workers equally. Both professors’ curves are solid lines, and the cafeteria workers’ curves are shown as dotted. In the case of profile $(A)$, the number of cafeteria workers in quarantine is around a hundred more compared with the number of professors in quarantine over the eighty-day period. In the case of profile $(B)$, the number of cafeteria workers in quarantine and the number of professors in quarantine remain practically the same until around day forty, where on average around fifty professors more are in quarantine compared to cafeteria workers. As expected, the different allocation profiles lead to different numbers of professors and cafeteria workers being in quarantine at any one point in time. This simple example shows the flexibility offered by our approach to find solutions that can be tailored according to the priorities of each educational institution.

\section{Reflections on Bridging Research and Practice}
\label{sec:reflections}

We briefly reflect on our experiences implementing practical tools for administrators at Mexican universities, and on our time building an international research collaboration with universities and research institutes within Mexico. A critical success factor to our work was establishing a solid and fluid working relationship with specific local councils belonging to the National Network of State Councils of Science and Technology (\href{https://www.rednacecyt.org/conocenos}{REDNACECYT})\footnote{In the case of Mexico, the national umbrella organisation supporting scientific research and development is the National Council of Science and Technology (CONACyT). In parallel, each state has an established local research council from REDNACECYT. These councils fund research and serve as a liason between the research community, practitioners, policymakers, industry and other external stakeholders. }.  This allowed us to establish collaborations with on-site facilities such as biology labs. Moreover, the expertise of local researchers gave us insights into the local context, including how to navigate regional administrative landscapes, which directly fed into the project specification and minimised chances of failure. In our case, we also needed to establish trust with the health experts and university administrators who ultimately would be using our tools. Our connections with local partners helped us maintain a feedback loop required in order to make the tool as useful as possible while reducing unnecessary complexity. In particular, administrator feedback allowed us to implement the bucketing technique described above, which drastically reduces the number of solutions to consider.

In addition, trust was particularly important with respect to privacy and data ownership, especially when dealing with sensitive health information. Here we highlight the benefits of working with aggregate instead of individual-level information, since the use of the latter not only raises questions around privacy and security, but also risks incurring significant delays due to compliance with data protection regulations. Instead, our algorithms were designed from the ground up to work only with aggregate information, and all calculations to obtain these aggregates from personal information were performed locally by the university. This ensured that sensitive data remained safe with the university, without compromising the efficiency of our strategy.

We also note that our team members' multi-lingual backgrounds greatly helped with communications from an early stage. Moreover, three team members are based in Mexico, which provides local insights and improves communication and coordination with our partner university. Our core team is dispersed in multiple countries and cities, and we believe that our remote-first model of collaboration will continue to be fruitful.


\section{Closing Remarks}
\label{sec:future-work}

\paragraph{Future work.}
In the immediate future, we look forward to assessing the performance of our pilot. As test results are fundamentally limited, we will not be able to precisely ascertain who within the population is infected, but we will be able to observe test results from the solutions chosen by university administrators, as well as the number of individuals told to self-isolate through our containment protocol. Furthermore, we will also assess how the populations from different campuses perceive the algorithmically-assisted decision-making process for test allocations, and comply with the subsequent containment procedure. In addition, we will increase the scope of the pilot to include other universities within Mexico.

Looking forward, we wish to explore testing resource allocation paradigms with richer classes of testing and containment mechanisms. Our current mechanism operates in the regime where an institution resumes regular activity and testing resources are utilised to preemptively prevent asymptomatic individuals from spreading the virus and disrupting these normal activities. When infection rates are sufficiently high, policy makers may instead choose to implement a lockdown in order to eliminate viral spread. In this regime, group testing can instead be used to search for the healthy among those in lockdown, rather than searching for the infected amongst an unconstrained population. In this case, we can formulate a similar optimisation problem for maximising the number of healthy individuals found in a heterogeneous population. Indeed, our testing and containment strategy from Section \ref{sec:model} can be augmented to allow the mechanism to categorically place individuals in self-isolation without them having been in a positive test (akin to placing an entire population category in lockdown).

Finally, a research direction we are currently exploring is that of applying reinforcement learning techniques directly in our testing and containment paradigm. We consider policies that have knowledge of the heterogeneity within a population, as well as a limited testing budget to allocate across the population, to inform containment decisions. Crucially, the policy will not be memory-less, which we hope will provide insight into non-trivial allocation mechanisms over a longer time horizon. 

\paragraph{Conclusion.}
The optimisation-based approach presented here allocates limited \covid{} tests within an educational institution. We note that our approach can be applied to any context in which a heterogeneous population can be meaningfully partitioned into categories. Furthermore, the web application created in collaboration with our partners in Mexico allows university administrators to intuitively visualise the trade-offs between different Pareto-dominant testing strategies, facilitating their decision-making process. As we are implementing our methodology with partner institutions in Mexico, we hope that these techniques can be of greater social good in geographies where testing resources are limited.  

\paragraph{Acknowledgements.}
We would like to thank the ACM Special Interest Group on Economics and Computation (\href{https://www.sigecom.org/}{SIGecom}) for their generous support of this work through a Global Challenges in Economics and Computation grant (\href{https://www.gcec.org/rfp}{GCEC '20}). We also thank Rubén López-Revilla, Salvador Ruiz-Correa, Angel Alpuche Solis from the Potosinian Institute of Scientific Research and Technology (\href{https://www.ipicyt.edu.mx/}{IPICYT}) and Rosalba Medina Rivera from the San Luis Potosí Council of Science and Technology (\href{https://slp.gob.mx/copocyt/Paginas/Inicio.aspx}{COPOCYT}) for their insight and support in applying our work in Mexico. We would also like to thank Martín De la Cruz, Ana Paula Valdez Hernandez, Alicia Minerva Ortíz Rojas, Herbert González Esqueda, Paulina Campos and Roberto Ponce-Lopez from the Instituto Tecnologico y de Estudios Superiores de Monterrey (\href{https://tec.mx/en}{ITESM}) for their crucial support in running our pilot within their campuses. Finally, we also thank Paul Klemperer, Christopher Bronk Ramsey and Divya Sridhar for key initial discussions.

\newpage
\bibliographystyle{abbrv}
\bibliography{library}

\appendix

\section{The Web Application}
\label{sec:appendix_web}
\begin{figure}[ht]
    \centering
    \includegraphics[width=0.95\columnwidth]{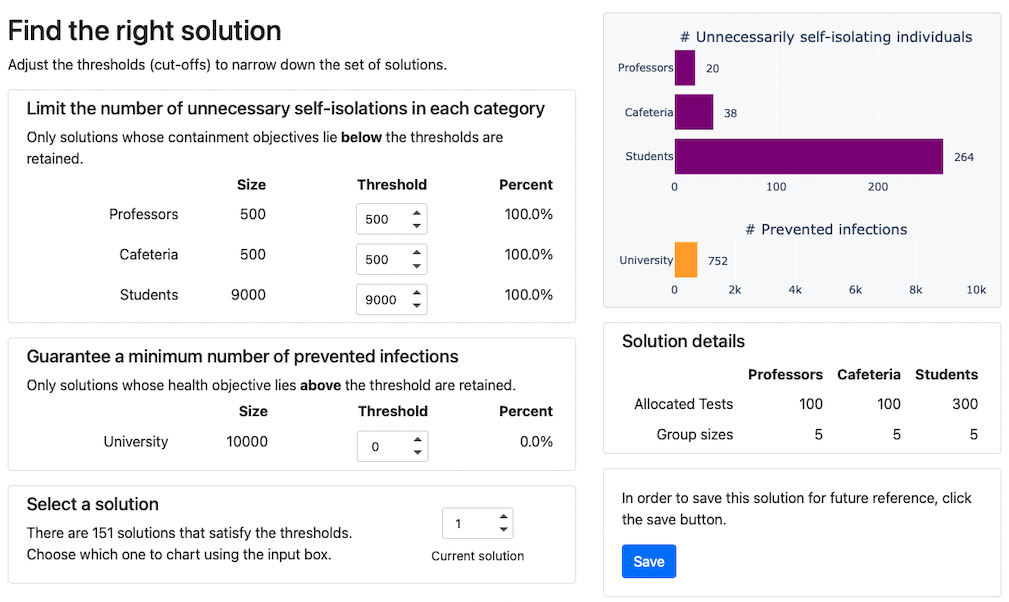}
    \caption{Our web application allows university administrators to explore the outcomes of different Pareto-optimal testing strategies.}
    \label{fig:webapp}
\end{figure}

A demo of our web application will be available at \href{http://eaamo-demo.azurewebsites.net}{our website}. Moreover, the source code is provided with instructions under an open source licence at \href{https://github.com/hguzmang/eaamo}{our GitHub repository}. Figure~\ref{fig:webapp} shows a screenshot of our application.

\section{Modelling the Health Objective}
\label{sec:appendix_health_objective}
Recall that $O_H(\tb, \gb)$ denotes the expected number of critical cases that are prevented in our single-step contagion model under testing strategy $(\tb, \gb)$. In order to state the expected number of critical cases that occur for a given testing strategy, we first determine the probability that an individual who is healthy and not self-isolating is infected by one of its neighbours in the population graph.

Suppose $w$ is an individual from $C_i$ who is healthy and not self-isolating following testing (but prior to the contagion step). We first study the probability that $w$ is infected by someone from category $C_j$, where $1 \leq j \leq k$.
As $t_j$ groups of size $g_j$ are chosen for testing uniformly at random from $C_j$, $n_j - t_j g_j$ individuals in $C_j$ remain untested. It follows that the number of untested contacts of $w$ in $C_j$ can be modelled as a random variable $\ub \sim \mathsf{Bin} \left( d_{ij}, \frac{n_j - t_jg_j}{n_j} \right)$. Moreover, each untested contact has an i.i.d.~probability $p_j$ of being infected. We let $\lb_j \sim \mathsf{Bin}(\ub, p_j)$ denote the number of infected contacts of $w$ in $C_j$. Putting these together, we get that $\lb_j \sim \mathsf{Bin} \left( d_{ij}, \frac{p_j (n_j - t_jg_j)}{n_j} \right)$. 

This allows us to determine the probability that $w$ is not infected by any of its $\lb_j$ infected contacts in $C_j$. Since an infected individual fails to infect a healthy contact with probability $1 - \pi_{ij}$, the probability that $w$ is not infected by any of their contacts in $C_j$ is given by $(1-\pi_{ij})^{\lb_j}$. In order for $w$ to remain healthy in our one-step contagion model, it must avoid infections from contacts across all categories. Thus, the overall probability of remaining healthy is
$
\prod_{j=1}^k(1-\pi_{ij})^{\lb_j}
$
and it follows that $w$ is infected with probability $1 - \prod_{j=1}^k(1-\pi_{ij})^{\lb_j}$.


We now analyse the number of critical cases that occur following our testing and containment mechanism under testing strategy $(\tb, \gb)$. This number can be compared to the outcome when no tests are applied (which can be understood as testing strategy $(\bm{0}, \bm{0})$) to obtain the number of critical cases prevented by strategy $(\tb, \gb)$.

Let $z_i$ denote the probability that an individual $w$ from $C_i$ is healthy and not self-isolating following testing (but prior to the contagion step). Note that this can only happen if $w$ is either not part of any group tests, or they are tested and the outcome is negative. These disjoint events happen with probability $\frac{n_i - t_i g_i}{n_i} q_i$ and $\frac{t_i g_i}{n_i}q_i ^ {g_i}$, respectively. (Recall that $q_i = 1 - p_i$.). It follows that $z_i = \frac{n_i - t_i g_i}{n_i}q_i + \frac{t_i g_i}{n_i} q_i ^ {g_i}$.

Next recall that a healthy individual $w$ is infected by any of its contacts in the contagion step with probability $1 - \prod_{j=1}^k(1-\pi_{ij})^{\lb_j}$, where we once more let 

\[\lb_j \sim \mathsf{Bin} \left( d_{ij}, \frac{p_j (n_j - t_jg_j)}{n_j} \right) \]

denote the number of infected contacts of $w$ in category $C_j$.
Moreover, if $w$ becomes infected in the contagion step, they have a $v_i$ probability of becoming critical.
This in turn means that the probability that an individual in $C_i$ is initially healthy and then develops a critical infection from contagion is $v_i z_i \left ( 1-\prod_{j=1}^k \left (1-\pi_{ij} \right )^{\lb_j} \right )$.
Hence the total number of critical infections from contagion is given by the random variable
\[
\sum_{i=1}^k n_i v_i z_i \left(1 - \prod_{j=1}^k(1-\pi_{ij})^{\lb_j}\right).
\]
By applying linearity of expectation and exploiting the independence of random variables $\lb_j$ we obtain its expectation
\begin{equation}
\label{eq:expected-critical-infections}
\sum_{i=1}^k n_i v_i z_i \left ( 1 - \prod_{j=1}^k \mathbb{E} \left [ (1-\pi_{ij})^{\lb_j} \right ] \right ).
\end{equation}
It remains to compute $\mathbb{E} \left[ (1-\pi_{ij})^{\lb_j} \right]$. We can transform this expression into $\mathbb{E} \left [ \exp (\ln(1-\pi_{ij})){\lb_j}) \right]$, and substitute $t = \ln(1-\pi_{ij})$ into the closed form expression of the moment-generating function%
\footnote{The moment-generating function for random variable $X \sim \mathsf{Bin}(n,p)$ is given by $M_{X}(t)=(pe^t + 1-p)^n$.}
for a binomially distributed random variable to obtain
\[
\mathbb{E} \left[ (1-\pi_{ij})^{\lb_j} \right] = \left (1-\frac{\pi_{ij}p_j (n_j - t_jg_j)}{n_j} \right)^{d_{ij}}.
\]
Substituting this expression into \eqref{eq:expected-critical-infections}, we see that the expected number of total critical infections from contagion under testing strategy $(\tb, \gb)$ is given by
\begin{equation}
\label{eq:expected-total-criticals}
f_H(\tb, \gb) =
\sum_{i=1}^k n_i v_i z_i \left( 1 - \prod_{j=1}^k \left(1-\frac{\pi_{ij}p_j (n_j - t_jg_j)}{n_j}\right)^{d_{ij}} \right).
\end{equation}

Note that we can determine the number of critical cases that occur without any testing by evaluating \eqref{eq:expected-total-criticals} with testing strategy $(\bm{0}, \bm{0})$. (In this case $z_i$ is just $q_i$.) Putting this together, we define our healthcare objective as follows:
\begin{equation*}
O_H(\tb, \gb) = f_H(\bm{0},\bm{0}) - f_H(\tb, \gb)
\end{equation*}

\section{The Bucketing Scheme}
\label{sec:appendix_buckets}
Both in our simulations and in the scenarios utilising the data provided by our university partners from Mexico, the Pareto frontier computation outputs thousands of possible solutions, $|S_P| > 1000$. As it can be a daunting task for policy makers to select appropriate solutions for their means, we have developed a bucketing scheme that enables the end user to choose as many solutions as they want the algorithm to generate. The bucketing works in the following way: for each of the objectives $O_H$ and the $O_{Q,i}$'s we set bucket sizes for each, denoted by $\rho_H$ for the health objective and $\rho_{Q_i}$ for each of the quarantine objectives. For each of the objectives, we round up each to the nearest multiple of the bucket size,
$\lfloor \frac{O_{H}(\tb, \gb)}{\rho_{H}} \rceil\rho_{H}$ and $\lfloor \frac{O_{Q,i}(\tb, \gb)}{\rho_{Q_i}} \rceil\rho_{Q_i}$. We also redefine Pareto-dominance for two distinct testing and containment mechanisms $(\tb, \gb)$ and $(\tb ', \gb ')$ by stipulating that $(\tb, \gb)$ \emph{Pareto-dominates} $(\tb ', \gb ')$ if and only if the following hold:
\[\left \lfloor \frac{O_{H}(\tb, \gb)}{\rho_{H}} \right\rceil\rho_{H} \geq \left \lfloor \frac{O_{H}(\tb ', \gb ')}{\rho_{H}} \right \rceil\rho_{H} \text{ and }\] 

\[\left \lfloor \frac{O_{Q,i}(\tb, \gb)}{\rho_{Q,i}} \right \rceil \rho_{Q,i} \leq \left \lfloor \frac{O_{Q,i}(\tb ', \gb ')}{\rho_{Q,i}} \right \rceil \rho_{Q,i} \ \forall i \in [n].\]
To obtain the desired number of solutions we utilise binary search. We first calculate the ranges of values taken by the health objective and of each quarantine objectives without bucketing in the Pareto frontier. We denote these ranges by $R_H$ for the health objective and $R_{Q,i}$ for the $i$-th quarantine objective. Given these ranges, we set the initial buckets as $\bm{\rho} =\bm{R} = (R_H, R_{Q,1},\dots, R_{Q,k} )$. We then perform binary search over the auxiliary parameter $\alpha \in[0,1]$ in order to find an $\alpha$ value such that the bucket sizes given by $\bm{\rho} = \alpha \bm{R}$ result in approximately the desired number of solutions.

\end{document}